\documentclass[reqno]{amsart}

\usepackage{amssymb, latexsym, amsmath, amsfonts, amscd}
\usepackage{lscape,color}
\usepackage{mathrsfs, hyperref}
\usepackage{verbatim}
\usepackage{appendix}
\usepackage{tikz}
\usetikzlibrary{quantikz}
\usepackage{multirow}
\usepackage{graphicx}
\usepackage{braket}

{\theoremstyle{definition}}

\newcommand{\Z}{\mathbb{Z}}

\newcommand{\X}{\mathbf{X}}
\newcommand{\Y}{\mathbf{Y}}

\usepackage[matha, mathx]{mathabx}
\definecolor{blue}{rgb}{0,0,1}

\title[Eavesdropping on the BB84 Protocol]{Eavesdropping on the BB84 Protocol using Phase-Covariant Cloning: Experimental Results}

\author[B. Pigott]{Brian Pigott}
\address{Wofford College\\pigottbj@wofford.edu}

\author[E. Campolongo]{Elizabeth G. Campolongo}
\address{The Ohio State University\\ e.campolongo479@gmail.com}

\author[H. Routray]{Hardik Routray}
\address{Rutgers University\\ hardikroutray.physics@gmail.com}

\author[A. Khan]{Alex Khan}
\address{University of Maryland, QLab\\ askhan@umd.edu}

\date{\today}

\begin{document}

\begin{abstract}
	Though the BB84 protocol has provable security over a noiseless quantum channel, the security is not proven over current noisy technology. The level of tolerable error on such systems is still unclear, as is how much information about a raw key may be obtained by an eavesdropper.
	We develop a reproducible test to determine the security--or lack thereof--of the protocol in practice.
This enables us to obtain an experimental estimate of the information that can be obtained using asymmetric phase-covariant cloning to eavesdrop on the BB84 protocol.
\end{abstract}

\maketitle

\section{Introduction}

The history of hidden or disguised communication dates back to the early days of human records and civilization.
Despite a rich history, the only truly (provably) secure encryption scheme was developed in the early twentieth century--the \textit{one-time-pad} cipher \cite{kahn1996}. To achieve this ultimate security, all that is required is a simple substitution cipher with two key distinctions: (1) the encryption key must be a randomly generated string the same length of the message (or longer), and (2) the encryption key must only be used for a single message, then discarded. The latter requirement poses the greatest impediment to its implementation; it is not feasible to generate long random keys for single messages when each party must have the shared key. Modern systems have endeavored to achieve practical security while removing the key-sharing challenge through what's known as \textit{asymmetric encryption}. One of the most well-known (and commonly implemented) such systems is RSA, which relies on the intractability of large number factorization for security. For an overview of RSA and its vulnerabilities, see Chapters~10 and 17 of \cite{easttom2021} and the references therein. These systems allow for any two parties to communicate securely without ever exchanging keys privately using the asymmetry of the scheme. Each has their own public and private key, so that a party wishing to communicate can encrypt a message using their desired recipient's public key, and only that person will be able to decrypt it. For added message integrity, they may choose to add a layer with their own private key, so that only their public key can decrypt, further proving who sent the original message.

The advent of quantum computing provided a new avenue along which to tackle the number theory question of efficient prime factorization: Shor's Algorithm leverages quantum computing to factor large ($N$-digit) numbers into primes in polynomial ($\log N$) time, much more efficiently than the classical Euclidean Algorithm \cite{easttom2021}. 
In response to classical computing advances, the recommended digit length of RSA key prime numbers has increased for intractability of factoring attacks. So far, implementations of Shor's Algorithm on current hardware are limited, however, the undeniable reality is that quantum computing hardware \textit{will} be able to execute this algorithm at a sufficiently high fidelity to render RSA insecure. As a result, there is a push to move beyond these prime number based systems of encryption to more resilient encryption standards, such as the lattice based systems selected by the National Institute of Standards and Technology \cite{NIST}. However, even these systems could be rendered vulnerable to quantum attacks should an efficient quantum algorithm be developed to solve their underlying mathematical problems. 

Hence, we return to the only provably secure encryption system, the one-time-pad, which will not suffer the same fate from quantum algorithms. In fact, it is the advent of quantum computing which has brought this encryption scheme back to the table, rendering it potentially feasible for use at scale. This brings us to 1984 and the work of Bennett and Brassard \cite{bennett2020quantum}: the BB84 protocol.

The first quantum key distribution protocol was introduced in 1984 with the publication of the BB84 protocol by Bennett and Brassard \cite{bennett2020quantum}. Since then, many other protocols have been developed and studied; see for instance \cite{scarani2009security}, \cite{vanassche}, \cite{xu2020secure} and the references therein. That said, BB84 remains among the most studied of the quantum key distribution protocols: it is frequently analyzed in academic publications and has often been implemented  in commercial products (see the beginning of Chapter 10 in \cite{vanassche} and the references found there). We provide a detailed description of the BB84 protocol in Section \ref{section:bb84}, including the three classes of attacks: individual, collective, and general-coherent. In the current work we consider only individual attacks, wherein, 
under ideal conditions, the legitimate parties in the BB84 protocol (typically called Alice and Bob) can tolerate a qubit error rate of roughly $15\%$ while still being able to distill a secure key (see Section \ref{secret_key_rate_section} or \cite{vanassche} for further details).  In the current era of noisy quantum computers (the so-called NISQ era)  it is unclear how much error Alice and Bob can tolerate on such a device, nor is it clear how much information a potential eavesdropper (usually called Eve) might be able to obtain about a raw key.

This work is thus focused on estimating the amount of information that Eve is able to procure using an individual attack
 on current quantum hardware.  Specifically, we measure how much information an eavesdropper can obtain about a raw key when transmitted under the BB84 protocol using the optimal  eavesdropping approach (asymmetric phase-covariant cloning \cite{fuchs1997optimal}). 
 In the course of exploring this information bound we also determine the qubit error rate that is tolerable by Alice and Bob. In Section \ref{section:experimental_results} we describe an experiment simulating BB84 that we implemented on the quantum computer IonQ Harmony that aimed to uncover these quantities. To our knowledge, this is the first experimental result that estimates the information gained by an eavesdropper against the BB84 protocol.

The data gathered from our experiments present an interesting statistical problem. The central issue is to determine the points at which the qubit error rates of the legitimate and illegitimate parties agree. To do so we fit quadratic polynomials to the fidelity data obtained from our experiments and compute the points at which these curves intersect. Because these two curves are fit to experimental data, determining error bounds on the intersection points is more involved. This problem is solved in two ways: using a Monte-Carlo simulation and using a bootstrapping approach. Both of these approaches to this problem appear to be new. These techniques yield results that are found to be in good agreement with each other, see Section \ref{section:stats}.

\subsection{Organization.} In Section \ref{section:bb84} we introduce the BB84 protocol along with the notation that is used in the remainder of the paper. Section \ref{pccloning} reviews asymmetric phase-covariant cloning, the optimal strategy for eavesdropping on the BB84 protocol; this includes the implementation of the phase-covariant cloning that we used in our experiments which we describe in Section \ref{section:experimental_results}. The statistical analysis of the experimental data is provided in Section \ref{section:stats}, and our conclusions are presented in Section \ref{section:conclusion}.

\begin{comment}
The work of Bennett and Brassard \cite{bennett2020quantum} in 1984 produced the first quantum key distribution protocol and initiated a field of study that has exploded in the 21\textsuperscript{st} century.

We focus here on individual eavesdropping strategies, meaning that the eavesdropper, often referred to as Eve, chooses to probe each individual qubit. It is known (see \cite{fuchs1997optimal})  that the optimal such strategy is obtained using phase-covariant cloning. 

{\color{red}{Include the information bounds in the introduction and describe the approach that we take here? In other words, state that our goal is to investigate these bounds in the context of modern hardware.}}

We choose to work on the equator of the Bloch sphere (in the so-called equatorial bases), in part because this simplifies our cloning circuit.

Mention that NISQ era computers are equipped with noisy gates. Two-qubit gates are an order of magnitude worse than single qubit gates.
Cloning involves two-qubit gates so it's important to make measurements. 

The comment did not appear in the PDF, so missed this note here which should be added in to surrounding text.
\end{comment}

\medskip

\noindent{\textbf{Acknowledgements.}} The authors are grateful for the support of IonQ Harmony, the QuForce Innovation Fellowship (where this work was initiated), and for the guidance of Daniel Oi. The authors would also like to thank Dr. M. Cathey for suggesting the statistical techniques that were used in Section \ref{section:stats}.

\section{The BB84 Protocol}
\label{section:bb84}
The following notation is inspired by the monograph \cite{vanassche}, though it is modified to reflect our choice to work in the equatorial bases. This choice is informed by the circuit we use to implement 
phase-covariant cloning (see Figure \ref{basicCloneFigure}); it is phase covariant on equatorial qubits.
 For convenience we denote these bases by
\begin{equation*}
	\mathbf{X} = \{ \ket{+}, \ket{-} \} \qquad \text{and} \qquad \mathbf{Y} = \{ \ket{+i}, \ket{-i} \},
\end{equation*}
where
\begin{equation*}
	\ket{+} = \frac{1}{\sqrt{2}} \big ( \ket{0} + \ket{1} \big ), \qquad \ket{-} = \frac{1}{\sqrt{2}} \big ( \ket{0} - \ket{1} \big ),
\end{equation*}
and
\begin{equation*}
	\ket{+i} = \frac{1}{\sqrt{2}} \big ( \ket{0} + i \ket{1} \big ), \qquad \ket{-i} = \frac{1}{\sqrt{2}} \big ( \ket{0} - i \ket{1} \big ).
\end{equation*}
The BB84 protocol consists of a  sequence of three steps.

\medskip

\noindent{\textit{Step 1.}}
Let $\mathcal{X} = \{0,1\}$ and let $A \in \mathcal{X}$ be a random variable denoting Alice's key elements; Alice chooses these elements randomly and independently. In the standard BB84 protocol there are two rules for encoding the key which we denote by $u \in \{\mathbf{X}, \mathbf{Y} \}$. Alice randomly and independently chooses which encoding rule she uses for each key element.
\begin{itemize}
	\item If $u = \mathbf{X}$, then Alice prepares a qubit from the basis $\mathbf{X}$ as
	\begin{equation*}
		A \mapsto \frac{1}{\sqrt{2}} \big ( \ket{0} + (-1)^{A-1} \ket{1} \big ).
	\end{equation*}
	\item If $u = \mathbf{Y}$, then Alice prepares a qubit from the $\mathbf{Y}$ basis as
	\begin{equation*}
		A \mapsto \frac{1}{\sqrt{2}} \big ( \ket{0} + (-1)^{A-1} i \ket{1} \big ).
	\end{equation*}
\end{itemize}
As Alice encodes her key according to the rules described above, she transmits each corresponding qubit to Bob.

\medskip

\noindent{\textit{Step 2.}}
Upon receiving each qubit, Bob randomly chooses to measure in either the $\mathbf{X}$-basis or the $\mathbf{Y}$-basis, obtaining the result $B_{\X}$ or $B_{\Y}$. 

\medskip

\noindent{\textit{Step 3.}}
After sending a predetermined number of qubits, Alice reveals the encoding rule she use for each of them. Alice and Bob now sift their key, meaning that they discard the key elements in which Alice encoded using $u = \X$ (resp. $u = \Y$) and Bob measured in the $\Y$-basis (resp. the $\X$-basis). For the remaining (sifted) key elements, we denote Bob's measurements by $B$. 

Note that this part of the process happens on a public channel, meaning that an eavesdropper can be assumed to have knowledge of Alice's encoding rules.

\begin{figure}
	\centering
	\includegraphics[width=0.9\linewidth]{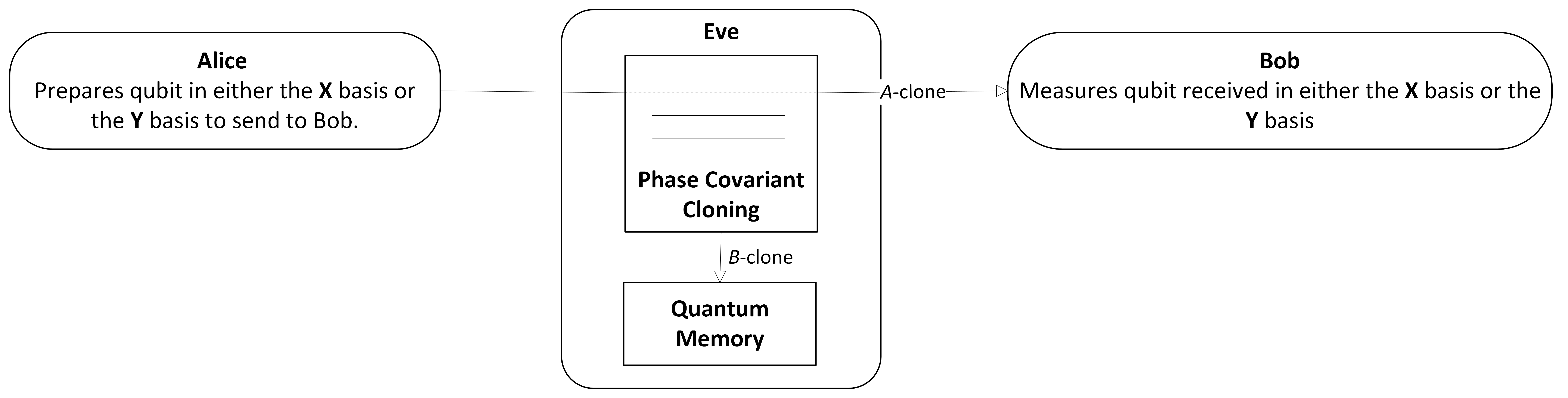}
	\caption[Eavesdropping on the BB84 protocol with phase-covariant cloning]{Eve uses asymmetric phase-covariant cloning to clone each individual qubit that Alice transmits to Bob. We assume that Eve transmits the $A$-clone to Bob while retaining the $B$-clone in her quantum memory until Alice and Bob reveal their measurement choices, at which point Eve measures each of her qubits individually.}
	\label{fig:drawing}
\end{figure}

\subsection{Eavesdropping}
The principal contribution of this work is an experimental estimate of the information that an eavesdropper, Eve, can obtain if Alice and Bob attempt to encode a key using the BB84 protocol.
Attacks on the BB84 protocol are grouped into three classes which we list here in order of increasing power: individual, collective, and general-coherent. For further details on the attacks we refer the reader to the survey article \cite{pirandola2020advances}, or to the monograph \cite{wolf2021quantum}. 

In an individual attack, Eve uses a fresh ancilla to interact with each qubit that Alice sends to Bob, and performs individual measurements on each of the output ancilla systems. We assume that Eve delays her measurements until the end of the protocol, after Alice and Bob have exchanged information about their basis choices. Alice and Bob can, in the case of an individual attack, tolerate a qubit error rate of roughly $15\%$ while still being able to distill a secure key, see \cite{fuchs1997optimal}.

In the case of a collective attack, Eve still uses a fresh ancilla to interact with each individual qubit that Alice sends to Bob, but the output of these ancillary systems are then stored in a quantum memory which is collectively measured at the end of the protocol after Alice and Bob have shared their basis choices. Under a collective attack, it is known that Alice and Bob are able to distill a secure key provided the qubit error rate is no higher than $11\%$, see \cite{pirandola2008symmetric}.

Under a general-coherent attack, Eve's ancillae and the qubits that Alice sends to Bob are subjected to a joint unitary interaction. Again, the ancillary output is stored in a quantum memory which is measured following the classical communication phase of the protocol. In the asymptotic scenario (i.e. when the number of signals $n \gg 1$ is extremely large, ideally infinite) general-coherent attacks can be reduced to a collective attack by using a random symmetrization routine that exploits the quantum de Finetti theorem \cite{renner2007symmetry, renner2008security, renner2009finetti}.

In the present work we focus on individual eavesdropping strategies. 
We do not impose restrictions on our eavesdropper (opting for the unbounded storage model) in recognition that technology is constantly evolving and one should not assume one's opponent is bounded by the same technological limitations. We work under the assumption that the number of signals $n \gg 1$ is very large (ideally, infinite), meaning that we need not consider finite-size effects. In this case it is known that the optimal eavesdropping strategy is for Eve to use asymmetric phase-covariant cloning as in Figure \ref{fig:drawing}, see \cite{fuchs1997optimal}.
Let $E$ be a random variable that contains any measurements that Eve makes. Thus $E$ contains everything that Eve has managed to infer from eavesdropping on the quantum channel.

\subsection{Secret Key Rate} \label{secret_key_rate_section}
The secret key rate that the legitimate parties can obtain with perfect reconciliation techniques is given by
\begin{equation}
	\label{secretkeyrate}
	S = \max \Big \{ I(A;B) - I(A;E), I(A;B) - I(B;E) \Big \},
\end{equation}
where $I(A;B)$ is the mutual information shared by the legitimate parties (Alice and Bob), and $I(A;E)$ (resp., $I(B;E)$) is the amount of information about Alice's key (resp., Bob's key) obtained by the eavesdropper. This secret key rate, \eqref{secretkeyrate}, can be obtained using one-way reconciliation \cite{devetak2005distillation}. Because the BB84 protocol is symmetric between Alice and Bob, we may assume that Alice's bits serve as a key. Thus, without loss of generality, we assume that Eve tries to obtain Alice's bits and that she tries to maximize $I(A;E)$.

It is known (see \cite{vanassche})  that the mutual information quantities $I(A;B)$ and $I(A;E)$ are given by
\begin{equation}
	\label{ABEinfo}
	I(A;B) = 1 - h(e_{B}) \qquad \text{and} \qquad I(A;E) = 1 - h ( e_{E} ),
\end{equation}
where $e_{B}$ is the error rate observed by Alice and Bob, $e_{E}$ is the error in the signal measured by Eve, and $h$ is the binary entropy for a binary distribution with probabilities $\{p, 1-p\}$:
\begin{equation*}
	h(p) = -p \log (p) - (1-p) \log(1-p).
\end{equation*}
At the endpoints where $p=0,1$ we define $h(0) = 0$ and $h(1) = 0$ for continuity.
In keeping with standard practice we note that the logarithm here is the logarithm with base $2$, i.e. $\log(\cdot) = \log_{2}(\cdot)$. 

Based on the quantities \eqref{ABEinfo} it transpires that Alice and Bob are able to derive a secure key provided $I(A;B) \geq I(A;E)$. In Section \ref{pccloning} we will see that when Eve uses phase-covariant cloning as her eavesdropping strategy, $e_{B}, e_{E} \leq 1/2$. As the binary entropy is increasing on the interval $(0,1/2)$ we thus find that $I(A;B) \geq I(A;E)$ provided $e_{B} \leq e_{E}$. In fact, 
the threshold  in the error rates occurs when $e_{B} = e_{E} = \frac{1}{2} - \frac{\sqrt{2}}{4}$; that is, Alice and Bob are able to distill a secure key provided $e_{B} < \frac{1}{2} - \frac{\sqrt{2}}{4}$. The corresponding theoretical bound on the mutual information that can be obtained by Eve is 
\begin{equation}
\label{Eve_mutual_info_bound}
I(A; E) \leq 0.39912.
\end{equation}

This theoretical bound on the information in terms of the disturbance induced by the eavesdropper was originally developed in \cite{fuchs1997optimal}. In \cite{Chiribella2005} the authors show that this bound can be achieved by the so-called phase-covariant cloning machines which we investigate in the next section.

\section{Asymmetric Phase-Covariant Cloning Machines}
\label{pccloning}
The no-cloning theorem prohibits one from producing a perfect copy (a clone) of an arbitrary quantum state \cite{wooters1982quantum}. However, it is possible to produce imperfect (approximate) copies of the state, as described in \cite{buvzek1996quantum}.
These \emph{universal quantum cloning machines} are designed so that the output fidelity of the copy is independent of the state that is meant to be cloned. In the case of the BB84 protocol, Eve needs only clone four states, each of which lie on the equator of the Bloch sphere. In restricting our cloning machine to the equatorial qubits, we obtain a so-called phase-covariant cloning machine, which also realizes an improvement in the output fidelity of the clones.

Our approach to phase-covariant quantum cloning machines is largely inspired by \cite{REZAKHANI2005278}. Here we restrict attention to the case of qubits, meaning that we take the dimension of our Hilbert spaces to be $d = 2$ in all cases. We refer the reader to \cite{REZAKHANI2005278} for the general case $d \geq 2$. Let $\mathcal{H}_{A}$ denote the Hilbert space of input states, let $\mathcal{H}_{B}$ denote the Hilbert space of the clone state, and let $\mathcal{H}_{X}$ denote the Hilbert space for the ancilla. Let $\{ \ket{i}_{A} \}_{i=0,1}$ be an orthonormal basis of $\mathcal{H}_{A}$, making similar definitions for $\mathcal{H}_{B},\mathcal{H}_{X}$. 

We briefly review the universal cloning machines of Buzek and Hillery \cite{BuzekHillery1998}. We assume that the ancilla is in some fixed initial state $\ket{\Sigma}$. Consider the transformation
\begin{equation}
	\label{uqcm}
	\ket{i}_{A} \ket{O}_{B} \ket{\Sigma}_{X} \mapsto \mu \ket{i}_{A} \ket{i}_{B} \ket{i}_{X} + \nu \sum_{\substack{0 \leq j \leq 1\\ j \neq i}} \Big ( \ket{i}_{A} \ket{j}_{B} + \ket{j}_{A} \ket{i}_{B} \Big ) \ket{j}_{X}.
\end{equation}
We first note that we can, without loss of generality, take the coefficients $\mu, \nu \in \mathbb{R}$.  We require that the transformation \eqref{uqcm} be unitary, universal\footnote{\textit{Universal} is used here to mean that the quality of the clone, as measured by the fidelity $F = \bra{\psi} \rho^{(\text{out})} \ket{\psi}$, is independent of the input state $\ket{\psi}$.}, symmetric ($\rho_{A}^{(\text{out)}} = \rho_{B}^{(\text{out})}$), and completely positive. In so doing one obtains the following relations:
\footnote{Note that we use the subscript $A(B)$ to refer to both $A$ and $B$ terms, eg., $\mathcal{H}_{A(B)}$ is meant to represent both Hilbert spaces $\mathcal{H}_A$ and $\mathcal{H}_B$.}
\begin{align*}
\rho_{A(B)}^{(\text{out})} &= \eta \rho_{A(B)}^{(\text{id})} + \frac{1 - \eta}{2} 1_{A(B)}\\
\mu^{2} = 2\mu \nu, &\qquad \mu^{2} = \frac{2}{3}, \qquad \nu^{2} = \frac{1}{6},\\
\eta &= \mu^{2},
\end{align*}
where $1_{A(B)}$ denotes the identity operator on $\mathcal{H}_{A(B)}$. The factor $\eta$ plays a particularly important role in what follows, in part because of its relationship to the fidelity $F$ of the clones:
\begin{equation*}
	F_{A(B)} = \frac{1+\eta}{2}.
\end{equation*}
The factor $0 <  \eta < 1$ is referred to as the \emph{shrinking factor}. From the equalities given above we see that the fidelity for a cloner of this sort is $F = 5/6$.

More generally one can define an \emph{asymmetric} cloning machine by breaking the symmetry present in the sum in \eqref{uqcm}:
\begin{equation}
	\label{assymetric_uqcm}
	\begin{aligned}
	\ket{i}_{A} \ket{O}_{B} \ket{\Sigma}_{X} \mapsto &\mu \ket{i}_{A} \ket{i}_{B} \ket{i}_{X} + \nu \sum_{\substack{0 \leq j \leq 1\\ j \neq i}} \ket{i}_{A} \ket{j}_{B} \ket{j}_{X}  + \xi \sum_{\substack{0 \leq j \leq 1\\ j \neq i}} \ket{j}_{A} \ket{i}_{B} \ket{j}_{X}.
	\end{aligned}
\end{equation}
A simple calculation reveals that if this machine is evaluated on a state $\ket{\psi} = \alpha_{0} \ket{0} + \alpha_{1} \ket{1}$ in $\mathcal{H}_{A}$, the output is given by
\begin{equation*}
	\begin{aligned}
	\rho_{A}^{(\text{out)}} &= 2 \mu \nu \ket{\psi} \bra{\psi} + \xi^{2} 1_{A}  + (\mu^{2} + \nu^{2} - \xi^{2} - 2 \mu \nu) \Big ( \vert \alpha_{0} \vert^{2} \ket{0} \bra{0} + \vert \alpha_{1} \vert^{2} \ket{1} \bra{1} \Big )
	\end{aligned}
\end{equation*} 
while the corresponding output for the $B$-clone is given by
\begin{equation*}
	\rho_{B}^{(\text{out)}} = 2 \mu \xi \ket{\psi} \bra{\psi} + \nu^{2} 1_{B}  + (\mu^{2} + \xi^{2} - \nu^{2} - 2 \mu \xi) \Big ( \vert \alpha_{0} \vert^{2} \ket{0} \bra{0} + \vert \alpha_{1} \vert^{2} \ket{1} \bra{1} \Big ).
\end{equation*}
In order for the output fidelity $F_{A}$ to be independent of the input state we require that the cloner have the form
\begin{equation}
	\label{clone_form}
	\rho_{A}^{(\text{out)}} = \eta_{A} \rho_{A}^{(\text{id})} + \frac{1 - \eta_{A}}{2} 1_{A},
\end{equation}
whence
\begin{equation}
	\label{uqcm_A_formulae}
	\eta_{A} = 2 \mu \nu, \quad \mu^{2} + \nu^{2} + \xi^{2} = 1, \quad \frac{1 - \eta_{A}}{2} = \xi^{2}, \quad \mu^{2} + \nu^{2} - \xi^{2} - 2\mu \nu = 0.
\end{equation}
Similarly, by requiring that the fidelity of the $B$-clone be state-independent one obtains
\begin{equation}
	\label{uqcm_A_formulae}
\eta_{B} = 2 \mu \xi, \quad \mu^{2} + \nu^{2} + \xi^{2} = 1, \quad \frac{1 - \eta_{B}}{2} = \nu^{2}, \quad \mu^{2} + \xi^{2} - \nu^{2} - 2\mu \xi = 0.
\end{equation}

Alternatively, observe that if the initial state has the form 
\begin{equation}
	\ket{\psi} = \frac{1}{\sqrt{2}} \Big ( \ket{0} + e^{i\phi} \ket{1} \Big ),
\end{equation}
with $\phi \in [0, 2\pi]$, then 
\begin{equation*}
	\vert \alpha_{0} \vert^{2} \ket{0} \bra{0} + \vert \alpha_{1} \vert^{2} \ket{1} \bra{1} =\frac{1}{2} 1_{A},
\end{equation*}
so that
	\begin{align}
	\rho_{A}^{(\text{out})} &= 2 \mu \nu \ket{\psi} \bra{\psi} + \left ( \xi^{2} + \frac{\mu^{2} + \nu^{2} - \xi^{2} - 2 \mu \nu}{2} \right ) 1_{A}, \label{rhoAout}\\
	\rho_{A}^{(\text{out})} &= 2 \mu \xi \ket{\psi} \bra{\psi} + \left ( \nu^{2} + \frac{\mu^{2} + \xi^{2} - \nu^{2} - 2 \mu \xi}{2} \right ) 1_{B}. 
	\label{rhoBout}
	\end{align}
Identifying coefficients in \eqref{rhoAout} and \eqref{rhoBout} with terms in \eqref{clone_form}, we find that
\begin{equation}
	\label{pqcm_shrinking_factors}
	\eta_{A} = 2 \mu \nu \qquad \text{and} \qquad \eta_{B} = 2 \mu \xi.
\end{equation}
In particular we find that the fidelity of the clones is
\begin{equation}
	F_{A} = \frac{1+ 2 \mu \nu}{2} \qquad \text{and} \qquad F_{B} = \frac{1 + 2\mu \xi}{2}.
\end{equation}

\subsection{Optimization.} \label{clone_optimization}
Following \cite{REZAKHANI2005278}, we say that the clones are \emph{optimal} if, for a fixed fidelity $F_{A}$ of the $A$-clone, the fidelity $F_{B}$ of the $B$-clone is as large as possible. Hence, the problem of maximizing fidelities can be reduced to optimizing the corresponding shrinking factors $\eta_{A}$ and $\eta_{B}$. Thus, the optimization problem is: for a fixed value of $\eta_{A}$, determine the largest possible value of $\eta_{B}$. We view this as a constrained optimization problem with constraints given by
\begin{equation}
	\label{etaA_constraint}
	\eta_{A} = 2 \mu \nu
\end{equation}
together with the normalization constraint
\begin{equation}
	\label{normalization_constraint}
	\mu^{2} + \nu^{2} + \xi^{2} = 1.
\end{equation}
Here the goal is to maximize the function $\eta_{B} = 2 \mu \xi$ subject to the constraints \eqref{etaA_constraint} and \eqref{normalization_constraint}. A Lagrange multiplier argument (which is included in the Appendix \ref{appendix:optimization}) reveals that the optimal clones satisfy the circle relation
\begin{equation}
	\label{circle_relation}
	\eta_{A}^{2} + \eta_{B}^{2} = 1
\end{equation}
in the 2-dimensional case.

\subsection{Mutual information}
The circle relation \eqref{circle_relation} has consequences for the mutual information shared between the various parties. Since the fidelity of the clones is given by the formula $F_{A(B)} = (1 + \eta_{A(B)})/2$, the error rates observed by Bob (who receives the $A$ clone) and Eve (who retains the $B$ clone) are given by
\begin{equation}
	\label{eB_eE}
	e_{B} = \frac{1-\eta_{A}}{2} \qquad \text{ and } \qquad e_{E} = \frac{1 - \eta_{B}}{2},
\end{equation}
respectively.\footnote{Error rates in terms of fidelity are given by $e_{B(E)} = 1 - F_{A(B)}$.} Notice, in particular, that since the shrinking factors $0< \eta_{A(B)}<1$, the error rates are bounded above by $\frac{1}{2}$, i.e., $e_{B}, e_{E} <  \frac{1}{2}$.
This means that both the fidelities and the qubit error rates can be rewritten using the circle relation:
\begin{equation*}
	(2F_{A}-1)^{2} + (2F_{B}-1)^{2} = 1,
\end{equation*}
and
\begin{equation}
	\label{circle_relation_qber}
	(1 - 2e_{B})^{2} + (1 - 2e_{E})^{2} = 1.
\end{equation}
In the case of the error rates, if one knows the error rate for Bob, $e_{B}$, they are now able to infer the error rate for Eve by solving \eqref{circle_relation_qber} for $e_{E}$, finding that
\begin{equation}
	\label{qber_formula}
	e_{E} = \frac{1}{2} - \sqrt{e_{B}(1 - e_{B})}.
\end{equation}
Returning to the mutual information formulas \eqref{ABEinfo} we see that
\begin{equation}
	\label{ABEinfo_Bob_qber}
	I(A;B) = 1 - h(e_{B}) \qquad \text{and} \qquad I(A;E) = 1 - h \left ( \frac{1}{2} - \sqrt{e_{B}(1 - e_{B})} \right ).
\end{equation}
It is important to note that \eqref{ABEinfo_Bob_qber} is symmetric in its arguments; if one knew the error rate for Eve, then Bob's qubit error rate could be expressed in terms of $e_{E}$, as could the mutual information between the parties. However, in practical applications, only Alice and Bob have access to the error rate that is measured in Bob's signal, $e_{B}$, meaning that they have quantities enabling them to estimate the information gained by Eve. On the other hand, Eve does not have access to these measurements and must rely on her knowledge of the quality of the clones over which she has control. Thus, it is most relevant to focus on this formulation.

 As discussed in Section~\ref{secret_key_rate_section},
Alice and Bob are able to distill a secure key provided $I(A;B) \geq I(A;E)$. This can be reduced to an inequality in the qubit error rates for Bob and Eve, namely, the requirement that $e_{B} \leq e_{E}$. With the formula \eqref{qber_formula} in hand we can now solve this inequality for $e_{B}$, finding that
\begin{equation}
	\label{critical_error}
	e_{B} < \frac{1}{2} - \frac{\sqrt{2}}{4}.
\end{equation}


\begin{figure}[ht]
	\centering
	\begin{quantikz}
		& \qw & \qw & \qw & \qw & \qw & \ctrl{1} & \ctrl{2} & \targ{} & \targ & & \qw \\
		& \gate{R_{y}(2 \theta_{1})} & \ctrl{1} & \qw & \targ{} & \gate{R_{y}(2 \theta_{3})} & \targ{} & \qw & \ctrl{-1} & \qw &  \qw \\
		& \qw & \targ{} & \gate{R_{y}(2 \theta_{2})} & \ctrl{-1} & \qw & \qw & \targ{} & \qw & \ctrl{-2} & \qw
	\end{quantikz}
	\caption{Circuit implementing phase-covariant  cloning adapted from \cite{PhysRevA.56.3446}. The top wire represents Alice's qubit (to clone), while Eve retains control over the two remaining wires. The middle wire produces the clone (for Eve) and the bottom is the ancilla.}\label{basicCloneFigure}
\end{figure}
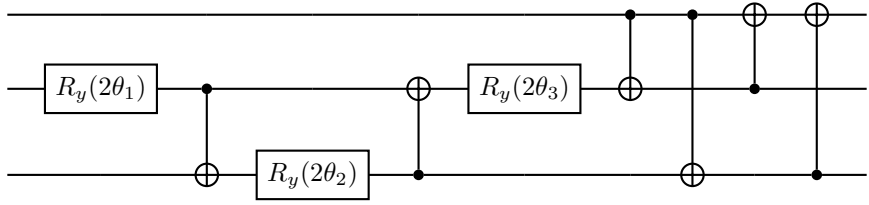

\subsection{Implementation of the Cloner}
To implement a phase-covariant cloning machine we use the circuit illustrated in Figure \ref{basicCloneFigure} which is taken from \cite{PhysRevA.56.3446}. Here the gate $R_{y}(\theta)$ has matrix representation
\begin{equation}
	\label{Rygate}
	R_{y}(\theta) = \left ( \begin{matrix} \cos \left ( \frac{\theta}{2} \right ) &  -\sin \left ( \frac{\theta}{2} \right )\\
	 \sin  \left ( \frac{\theta}{2} \right ) & \cos  \left ( \frac{\theta}{2} \right ) \end{matrix} \right )
\end{equation}
(in the computational basis). The top wire represents the qubit that we wish to clone. The remaining two wires are under the eavesdropper's control; the clone will be the output of the second (middle) wire while the bottom wire is an ancilla.

We assume that the input has the form of an equatorial qubit:
\begin{equation}
	\label{circuit_input}
	\ket{\psi}^{(\text{in})} = \frac{1}{\sqrt{2}} \Big ( \ket{0} + e^{i\phi} \ket{1} \Big ), \qquad \phi \in [0,2\pi).
\end{equation}
To facilitate our calculations we consider the cases of the $\ket{0}$ and $\ket{1}$ qubits separately.
In the case of the $\ket{0}$ qubit we find that
\begin{equation}
	\label{0qubit_output}
	\begin{aligned}
	\ket{0} \mapsto & \Big ( \cos(\theta_{1}) \cos(\theta_{2}) \cos(\theta_{3}) + \sin(\theta_{1}) \sin(\theta_{2}) \sin(\theta_{3}) \Big ) \ket{000}\\
	& + \Big (\cos(\theta_{1}) \cos(\theta_{2}) \sin(\theta_{3}) - \sin(\theta_{1}) \sin(\theta_{2}) \cos(\theta_{3}) \Big ) \ket{110}\\
	& + \Big ( \sin(\theta_{1}) \cos(\theta_{2}) \cos(\theta_{3}) - \cos(\theta_{1}) \sin(\theta_{2}) \sin(\theta_{3}) \Big ) \ket{101}\\
	& + \Big ( \cos(\theta_{1}) \sin(\theta_{2}) \cos(\theta_{3}) + \sin(\theta_{1}) \cos(\theta_{2}) \sin(\theta_{3}) \Big ) \ket{011},
	\end{aligned}
\end{equation}
and, similarly, for the $\ket{1}$ qubit we find that
\begin{equation}
	\label{1qubit_output}
	\begin{aligned}
		\ket{1} \mapsto &  \Big ( \cos(\theta_{1}) \cos(\theta_{2}) \cos(\theta_{3}) + \sin(\theta_{1}) \sin(\theta_{2}) \sin(\theta_{3}) \Big ) \ket{111}\\
		& + \Big (\cos(\theta_{1}) \cos(\theta_{2}) \sin(\theta_{3}) - \sin(\theta_{1}) \sin(\theta_{2}) \cos(\theta_{3}) \Big ) \ket{001}\\
		& + \Big ( \sin(\theta_{1}) \cos(\theta_{2}) \cos(\theta_{3}) - \cos(\theta_{1}) \sin(\theta_{2}) \sin(\theta_{3}) \Big ) \ket{010}\\
		& + \Big ( \cos(\theta_{1}) \sin(\theta_{2}) \cos(\theta_{3}) + \sin(\theta_{1}) \cos(\theta_{2}) \sin(\theta_{3}) \Big ) \ket{100}.
	\end{aligned}
\end{equation}

Recalling that our cloner has the form \eqref{assymetric_uqcm} and comparing coefficients with the cloner output above, we find that
\begin{equation}
	\label{coeffs_angles}
	\begin{aligned}
		\mu &= \cos(\theta_{1}) \cos(\theta_{2}) \cos(\theta_{3}) + \sin(\theta_{1}) \sin(\theta_{2}) \sin(\theta_{3})\\
		\nu &= \cos(\theta_{1}) \sin(\theta_{2}) \cos(\theta_{3}) + \sin(\theta_{1}) \cos(\theta_{2}) \sin(\theta_{3})\\
		\xi &= \sin(\theta_{1}) \cos(\theta_{2}) \cos(\theta_{3}) - \cos(\theta_{1}) \sin(\theta_{2}) \sin(\theta_{3}),
	\end{aligned}
\end{equation}
together with the requirement that
\begin{equation}
	\label{zero_condition}
	\cos(\theta_{1}) \cos(\theta_{2}) \sin(\theta_{3}) - \sin(\theta_{1}) \sin(\theta_{2}) \cos(\theta_{3}) = 0.
\end{equation}

\subsection{Selection of Angles} We now turn to the problem of determining values for the angles $\theta_{1}, \theta_{2}, \theta_{3}$ that determine an optimal phase-covariant cloner for equatorial qubits.

To begin we recall that the factors $\eta_{A}$ and $\eta_{B}$ are given by
\eqref{pqcm_shrinking_factors} and satisfy the circle identity \eqref{circle_relation}.

Using the identifications \eqref{coeffs_angles} together with the formulas \eqref{pqcm_shrinking_factors} we are able to express $\eta_{A}$ and $\eta_{B}$ in terms of $\theta_{1}, \theta_{2},$ and $\theta_{3}$. Indeed,
\begin{align*}
	\eta_{A} &= 
	2 \Big ( \cos^{2}(\theta_{1}) \sin(\theta_{2}) \cos(\theta_{2}) \cos^{2}(\theta_{3}) + \sin(\theta_{1}) \cos(\theta_{1}) \cos^{2}(\theta_{2}) \sin(\theta_{3})\cos(\theta_{3}) \\
	& \quad + \sin(\theta_{1}) \cos(\theta_{1}) \sin^{2}(\theta_{2}) \sin(\theta_{3}) \cos(\theta_{3}) + \sin^{2}(\theta_{1}) \sin(\theta_{2}) \cos(\theta_{2}) \sin^{2}(\theta_{3}) \Big )
\end{align*}
Observe that we can use the identity \eqref{zero_condition} to rewrite the two middle terms in this expression:
\begin{align*}
	\sin(\theta_{1}) \cos(\theta_{1}) \cos^{2}(\theta_{2}) \sin(\theta_{3})\cos(\theta_{3}) \!
	&= \sin(\theta_{1}) \cos(\theta_{2}) \cos(\theta_{3}) \!\Big ( \!\! \cos(\theta_{1}) \cos(\theta_{2}) \sin(\theta_{3}) \!\Big )\\
	&=\sin(\theta_{1}) \cos(\theta_{2}) \cos(\theta_{3}) \!\Big ( \!\! \sin(\theta_{1}) \sin(\theta_{2}) \cos(\theta_{3}) \!\Big )\\
	&= \sin^{2}(\theta_{1}) \sin(\theta_{2}) \cos(\theta_{2}) \cos^{2}(\theta_{3}),
	\intertext{and}
	\sin(\theta_{1}) \cos(\theta_{1}) \sin^{2}(\theta_{2}) \sin(\theta_{3}) \cos(\theta_{3}) \! &=
	\cos(\theta_{1}) \sin(\theta_{2}) \sin(\theta_{3}) \!\Big (\!\! \sin(\theta_{1}) \sin(\theta_{2}) \cos(\theta_{3}) \!\Big )\\
	&= \cos(\theta_{1}) \sin(\theta_{2}) \sin(\theta_{3}) \!\Big (\!\! \cos(\theta_{1}) \cos(\theta_{2}) \sin(\theta_{3}) \!\Big )\\
	&= \cos^{2}(\theta_{1}) \sin(\theta_{2}) \cos(\theta_{2}) \sin^{2}(\theta_{3}).
\end{align*}
Returning to our calculation of $\eta_{A}$, we have
\begin{align}
	\eta_{A} &= 2 \Big ( \cos^{2}(\theta_{1}) \sin(\theta_{2}) \cos(\theta_{2}) \cos^{2}(\theta_{3}) + \sin^{2}(\theta_{1}) \sin(\theta_{2}) \cos(\theta_{2}) \cos^{2}(\theta_{3}) \nonumber\\
	& \qquad + \cos^{2}(\theta_{1}) \sin(\theta_{2}) \cos(\theta_{2}) \sin^{2}(\theta_{3}) + \sin^{2}(\theta_{1}) \sin(\theta_{2}) \cos(\theta_{2}) \sin^{2}(\theta_{3}) \Big ) \nonumber\\
	&= 2 \Big ( \sin(\theta_{2}) \cos(\theta_{2}) \cos^{2}(\theta_{3}) + \sin(\theta_{2}) \cos(\theta_{2}) \sin^{2}(\theta_{3}) \Big ) \label{firstPythag}\\ 
	&= 2 \sin(\theta_{2}) \cos(\theta_{2}) \label{secondPythag} \\ 
	&= \sin(2 \theta_{2}). \label{sinSum}
\end{align}
Note that we made use of the Pythagorean identity, $\cos^{2}(2 \theta_{2}) + \sin^{2}(2 \theta_{2}) = 1$, in \eqref{firstPythag} and \eqref{secondPythag}, then simplified \eqref{sinSum} using the standard \textit{sine} summation formula.

In the case of the $\eta_{B}$ factor we proceed similarly, making use of \eqref{zero_condition} and previously noted trigonometric identities to obtain 
\begin{equation}
	\label{etaBinTrig}
	\eta_{B} =  \sin(2 \theta_{1}) \cos(2 \theta_{2}).
\end{equation}

In the case where the cloning transformation is optimal, $\eta_{A}$ and $\eta_{B}$ satisfy the circle relation \eqref{circle_relation}
In terms of our angles $\theta_{1}, \theta_{2}, \theta_{3}$, this now reads
\begin{equation*}
	\sin^{2}(2 \theta_{2}) + \sin^{2}(2 \theta_{1}) \cos^{2}(2 \theta_{2}) = 1.
\end{equation*}
Once again making use of the Pythagorean identity, we find that
\begin{equation*}
	\sin^{2}(2 \theta_{1}) = 1 \qquad \text{or} \qquad \cos^{2}(2 \theta_{2}) = 0. 
\end{equation*}
Notice that if $\cos(2 \theta_{2}) = 0$, then $\eta_{B} = 0$ and $\eta_{A} = 1$. While this presents itself as a possibility, it is not a particularly interesting one and we will not dwell on it, as this represents the case where the top qubit is not cloned. Of greater interest is the other restriction which requires $\sin(2\theta_{1}) = \pm 1$.
\begin{itemize}
	\item If $\sin(2 \theta_{1}) = 1$, then $\theta_{1} = \frac{n\pi}{4}$, for any $n = 4\ell + 1$, $\ell \in \mathbb{Z}$. 
	Returning this to \eqref{zero_condition} yields
	\begin{equation*}
		\cos(\theta_{2}) \sin(\theta_{3}) - \sin(\theta_{2}) \cos(\theta_{3}) = 0.
	\end{equation*}
	That is, we require that $\sin(\theta_{3} - \theta_{2}) = 0$, meaning that $\theta_{3} - \theta_{2} = k\pi$ for some $k \in \Z$. In what follows we will take $k = 0$, yielding $\theta_{2} = \theta_{3}$. This choice is also consistent with the requirement that
	\begin{equation*}
		\mu = \cos(\theta_{1}) \cos(\theta_{2}) \cos(\theta_{3}) + \sin(\theta_{1}) \sin(\theta_{2}) \sin(\theta_{3}) = \frac{1}{\sqrt{2}},
	\end{equation*}
	which emerges as a requirement of the optimal transformation.
	\item If $\sin(2\theta_{1}) = -1$, then $\theta_{1} = \frac{n\pi}{4}$, for any $n = 4\ell + 3$, $\ell \in \mathbb{Z}$. 
	Returning this to \eqref{zero_condition} leaves us with
	\begin{equation*}
		-\cos(\theta_{2}) \sin(\theta_{3}) - \sin(\theta_{2}) \cos(\theta_{3}) = 0,
	\end{equation*}
	which simplifies to $\sin(\theta_{3} + \theta_{2}) = 0$, meaning that $\theta_{2} + \theta_{3} = k\pi$ for some $k \in \Z$. Although this yields an equally valid equatorial phase-covariant cloning machine, we do not consider it further.
\end{itemize}

\medskip

\noindent{\textbf{Summary.}} The circuit presented at the beginning of this section implements a phase-covariant cloning machine if the angles $\theta_{1}, \theta_{2}, \theta_{3}$ are selected so that
\begin{equation}
	\label{angle_selection}
	\theta_{1} = \frac{\pi}{4}, \qquad \theta_{2} = \theta_{3}.
\end{equation}
From the above calculations and \eqref{eB_eE}, we conclude that the shrinking factors are given by
\begin{equation*}
\eta_{A} = \sin(2\theta_{2}) \qquad \text{and} \qquad \eta_{B} = \cos(2 \theta_{2}).
\end{equation*}
Recalling that these factors must be positive, the angle $\theta_{2}$ is thus restricted to the interval $(0, \frac{\pi}{4})$.

\subsection{Mutual Information under the Implementation}
Under the angle choice given in \eqref{angle_selection}, the coefficients of the cloner listed in \eqref{coeffs_angles} are as follows:
\begin{equation}
	\begin{aligned}
		\mu &= \frac{\sqrt{2}}{2} \\ 
		\nu &= \frac{\sqrt{2}}{2} \sin(2 \theta)\\
		\xi &= \frac{\sqrt{2}}{2} \cos(2 \theta),
	\end{aligned}
\end{equation}
where we have dropped the subscript notation for the angles, writing $\theta = \theta_{2} = \theta_{3}$.  In particular we find that the shrinking factors are given by
\begin{equation*}
	\eta_{A} = \sin(2\theta) \qquad \text{and} \qquad \eta_{B} = \cos(2 \theta).
\end{equation*}
Thus the fidelities of the clones are given by 
\begin{equation}
	\label{implementation_fidelities}
	F_{A}(\theta) = \frac{1 + \sin(2\theta)}{2} \qquad \text{and} \qquad F_{B}(\theta) = \frac{1 + \cos(2 \theta)}{2}.
\end{equation}

Recall from \eqref{eB_eE} that Bob's error rate is given by
\begin{equation*}
	e_{B}(\theta) = \frac{1}{2} \Big(1 - \eta_{A}(\theta) \Big) =  \frac{1}{2} - \frac{1}{2} \sin(2 \theta).
\end{equation*} 
Similarly, we find that Eve's error rate is given by
\begin{equation*}
	e_{E}(\theta) = \frac{1}{2} \Big(1 - \eta_{B}(\theta) \Big) = \frac{1}{2} - \frac{1}{2} \cos(2\theta).
\end{equation*}
In particular, using \eqref{ABEinfo}, the mutual information between Alice and Bob can now be written in terms of the angle $\theta$:
\begin{equation}
\label{IABtheory}
	I(A;B) = 1-h(e_B) = 1 - h \left ( \frac{1}{2} - \frac{1}{2} \sin(2 \theta) \right ).
\end{equation}
Similarly the mutual information between Alice and Eve is given by
\begin{equation}
\label{IAEtheory}
	I(A;E) = 1-h(e_E) = 1 - h \left ( \frac{1}{2} - \frac{1}{2} \cos(2 \theta) \right ).
\end{equation}
Recall from \eqref{critical_error} that Alice and Bob are able to distill a secure key if $e_{B} < \frac{1}{2} - \frac{\sqrt{2}}{4}$. We find that the angle at which this error rate is achieved is $\theta = \pi/8$. Thus, Bob's error rate, $e_B < \frac{1}{2} - \frac{\sqrt{2}}{4}$ when $\frac{\pi}{8} < \theta < \frac{\pi}{4}$, meaning that Alice and Bob will be able to distill a secure key for angles in this range.  

\section{Experimental Results}
\label{section:experimental_results}

In this section we describe our simulation of the BB84 protocol on IonQ Harmoy.
Using the circuit implementation given in Figure \ref{basicCloneFigure} we consider each of the elements of the $\mathbf{X}$ and $\mathbf{Y}$ bases. As the qubits sent by Alice in the BB84 protocol are independent of each other we treat each basis element separately. 

The experimental results obtained here were gathered as follows: for each of the BB84 states ($\ket{+}, \ket{-}, \ket{+i}, \ket{-i}$) we randomly (and uniformly) selected 100 values of the cloning angle $\theta = \theta_{2}$ from the interval $[0, \pi/4]$. Upon preparing the state to be cloned, we ran the circuit from Figure \ref{basicCloneFigure} for 100 shots with each of the randomly selected cloning angles. The fidelity of the clones was then computed and recorded. The statistical analysis of the measurements is presented in the next section.

\section{Statistical Analysis}
\label{section:stats}

Our goal is to estimate the amount of information gained by Eve using the cloning circuit in Figure \ref{basicCloneFigure}. The essential component in this is to determine the cloning angle at which the fidelity curves for Bob and Eve intersect, and to determine the qubit error rate for Alice and Bob at this angle. 

\begin{comment}
There are three separate quantities that deserve consideration:
\begin{enumerate}
	\item \textit{Information estimates available to Alice and Bob.} Here we recall \eqref{ABEinfo_Bob_qber}; the information available to both the legitimate parties and the illegitimate party can be calculated in terms of the error rate in Bob's qubits. Since Alice and Bob have access to this error rate, they can estimate $I(A;E)$ and $I(A;B)$.
	\item \textit{Information estimates available to Eve.} Eve doesn't have have access to any information other than the cloning angle that she has used in her eavesdropping. This means that she is forced to approximate the amount of information she obtains using theoretical values given in \eqref{IABtheory} and \eqref{IAEtheory}.
	\item \textit{Information estimates available to none of the parties.} Since we have access to the fidelities obtained in the eavesdropping process for all the parties involved, we are able to approximate the error rate that Bob and Alice must exceed to gaurantee their ability to distill a secure key.
\end{enumerate}
\end{comment}

\begin{figure}[ht]
	\centering
	\includegraphics[width=1\linewidth]{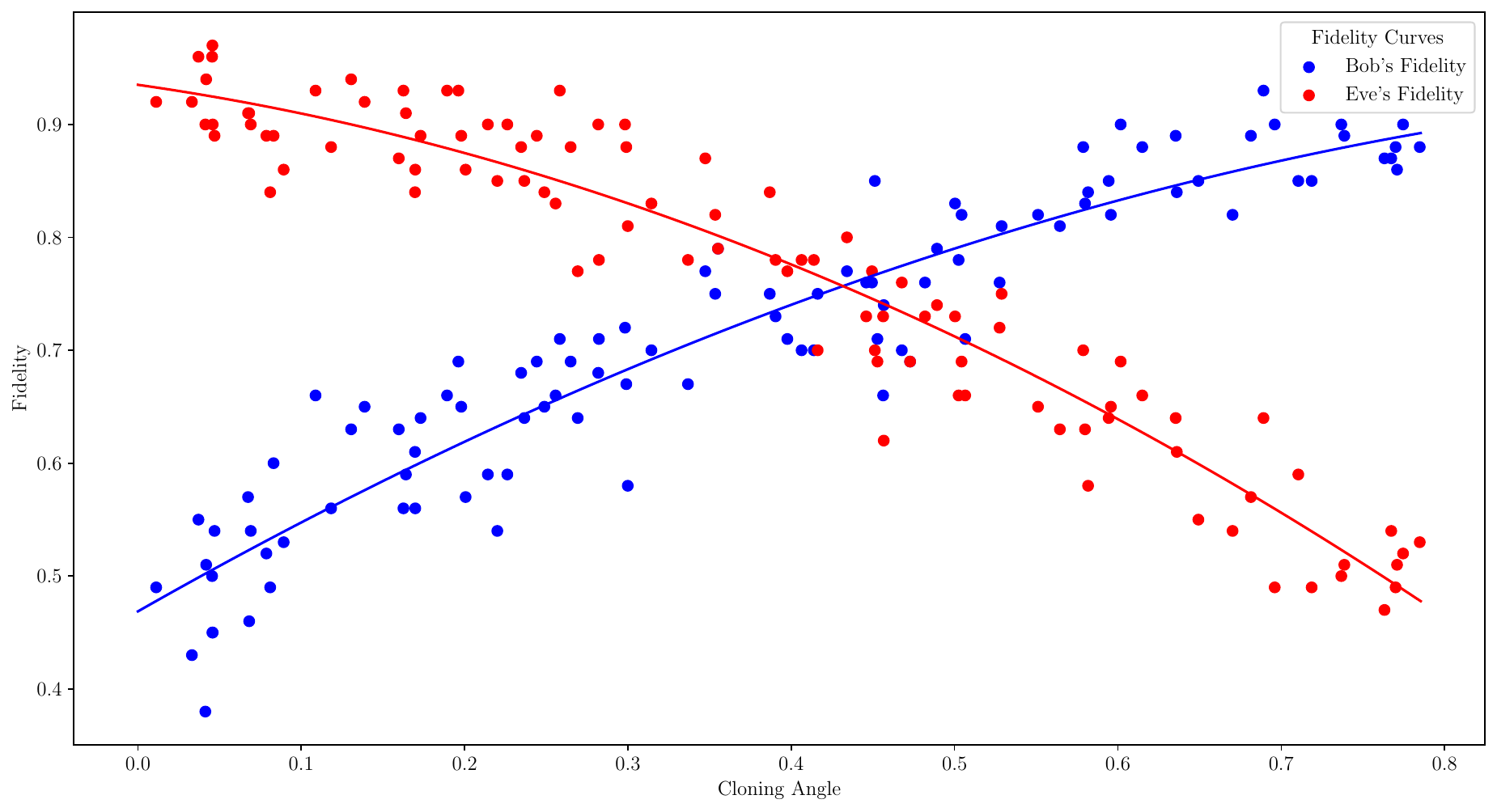}
	\caption{Scatter plot of the fidelity data obtained from the experiment conducted on the $\ket{-}$ qubit. The curves in this plot were obtained by fitting a quadratic polynomial to the experimental data. Recall that $F_{A(B)} = 1 - e_{B(E)}(\theta)$, so error rates decrease with increased fidelity.}
	\label{fig:fidelityplot}
\end{figure}

For each of the four BB84 states we obtained the following data: we have 100 randomly (and uniformly) selected cloning angles and  two fidelity measurements for each such angle: one for the fidelity of the clone that goes to Bob (the top wire in Figure \ref{basicCloneFigure}) and one fidelity for the clone retained by Eve (the middle wire in Figure \ref{basicCloneFigure}). A scatter plot of this data in the case of the $\ket{-}$ qubit is presented in Figure \ref{fig:fidelityplot}
; the corresponding scatter plots for the other qubits are similar. 

A quadratic polynomial was fit to each of the fidelity plots,\footnote{We opted for a quadratic polynomial for the fit as higher degree polynomials were found to overfit the data.} as shown in Figure~\ref{fig:fidelityplot}.
With two fidelity curves in hand (one for Bob's fidelity and one for Eve's fidelity) we determined their point of intersection, giving an estimate of the cloning angle and the qubit error rate; values of $\theta$ smaller than this intersection point yield better fidelities for Eve compared with those of Bob, meaning that Alice and Bob are unable to distill a secret key in this regime. Of key interest are the fidelities at the intersection of the curves; Alice and Bob are able to distill a secure key for corresponding qubit error rates lower than this experimentally determined rate. We recall that the theoretical qubit error rate that is tolerable for Alice and Bob is roughly 15$\%$.

To determine the confidence intervals for these measurements, we used two different techniques: a Monte-Carlo analysis based on errors in the coefficients, and a bootstrapping approach that is based on the data itself. These approaches and their results are presented below.

\subsection{Monte-Carlo Analysis}
One can reasonably assume that the coefficients in the quadratic polynomials that are fit to the fidelity data are normally distributed with standard deviation, $s$, given by the square root of the diagonal entries of the covariance matrix and with mean $\mu$, the value of the coefficient. In this analysis we randomly select coefficients from the normal distribution with the given mean and standard deviation and form the associated quadratic polynomial. This procedure is carried out on both the fidelity curves coming from each of the clones. In this way we obtain two new curves, each with randomly selected coefficients. We again compute the intersection points of these curves. This procedure was repeated ten thousand times. From the intersection data obtained in this way we are able to determine a mean error rate and an associated confidence interval. The results are presented in Tables \ref{MonteCarlo_angle} and \ref{MonteCarlo_fidelity}.

\begin{table}[ht]
	\centering
	\begin{tabular}{| c | c | c |}
	\hline
	Basis Element & Mean & 95$\%$ Confidence Interval\\
	\hline
		$\ket{-}$ & 0.43797 & (0.43697, 0.43897 )  \\
	\hline
		$\ket{+}$ & 0.411365 & (0.41029, 0.42442) \\
		\hline
		$\ket{-i}$ & 0.447044 & (0.44568, 0.44840) \\
		\hline
		$\ket{+i}$ & 0.41685 & (0.41575, 0.41795) \\
		\hline
\end{tabular}
\medskip
\caption{Statistics from Monte-Carlo simulation for angle of intersection.}
	\label{MonteCarlo_angle}
\end{table}

\begin{table}[ht]
	\centering
	\begin{tabular}{| c | c | c |}
		\hline
		Basis Element & Mean & 95$\%$ Confidence Interval\\
		\hline
		$\ket{-}$ & 0.24466  & (0.24409, 0.24522)  \\
		\hline
		$\ket{+}$ & 0.26845 & (0.26783, 0.26908) \\
		\hline
		$\ket{-i}$ & 0.18670 & (0.18604, 0.18735) \\
		\hline
		$\ket{+i}$ & 0.17996 & (0.17949, 0.18044) \\
		\hline
	\end{tabular}
\medskip
	\caption{Statistics from Monte-Carlo simulation for the qubit error rate at the intersection point.}
	\label{MonteCarlo_fidelity}
\end{table}

\subsection{Bootstrap Approach}

In the bootstrap approach we return to the fidelity data obtained during the experiment. From the one hundred data points we randomly select one hundred points, with replacement.  We again fit curves (quadratic polynomials) to the resulting newly  obtained data sets and determine their points of intersection. The resulting data provides a sample distribution for the intersection points from which we are able to determine a mean and a corresponding confidence interval. The result are summarized in Tables \ref{Bootstrap_angle} and \ref{Bootstrap_fidelity}.

\begin{table}[ht]
	\centering
	\begin{tabular}{| c | c | c |}
		\hline
		Basis Element & Mean & 95$\%$ Confidence Interval\\
		\hline
		$\ket{-}$ & 0.43183  & (0.43169, 0.43196)  \\
		\hline
		$\ket{+}$ &  0.40488 & (0.40471, 0.40504) \\
		\hline
		$\ket{-i}$ & 0.44004 & (0.43986, 0.44021) \\
		\hline
		$\ket{+i}$ & 0.41029 & (0.41014, 0.41044) \\
		\hline
	\end{tabular}
\medskip
	\caption{Statistics from the bootstrap simulation  for angle of intersection..}
	\label{Bootstrap_angle}
\end{table}

\begin{table}[ht]
	\centering
	\begin{tabular}{| c | c | c |}
		\hline
		Basis Element & Mean & 95$\%$ Confidence Interval\\
		\hline
		$\ket{-}$ & 0.24318  & (0.24307, 0.24328)  \\
		\hline
		$\ket{+}$ & 0.26747 & (0.26738, 0.26758) \\
		\hline
		$\ket{-i}$ & 0.18430 & (0.18322, 0.18438) \\
		\hline
		$\ket{+i}$ & 0.17789 & (0.17782, 0.17795) \\
		\hline
	\end{tabular}
\medskip
	\caption{Statistics from bootstrap simulation for fidelity at the intersection point.}
	\label{Bootstrap_fidelity}
\end{table}

\section{Conclusion}
\label{section:conclusion}
We note that the results of the statistical analyses presented in Section~\ref{section:stats} above (see Tables \ref{MonteCarlo_angle}, \ref{MonteCarlo_fidelity}, \ref{Bootstrap_angle}, and \ref{Bootstrap_fidelity}) agree to reasonably high accuracy. From our description of the BB84 protocol in Section~\ref{section:bb84},  statistically each of the relevant qubits will occur in a key $25\%$ of the time. In a sufficiently long sifted key, we would thus expect to find that each qubit has occurred with frequency $25\%$. From the statistics in Table \ref{Bootstrap_fidelity} (one could alternatively use the statistics from Table~\ref{MonteCarlo_fidelity}) we find that the cumulative error rate observed by Bob is $0.21821$ (the $95\%$ confidence interval for this mean is $(0.21803, 0.21839)$). Recall that the theoretical error bound that Alice and Bob can tolerate is $\frac{1}{2} - \frac{\sqrt{2}}{4} \approx 0.14645$, placing the experimental error bound roughly $7\%$ higher than theory predicts. This value is consistent with the error observed by IonQ Harmony: the error rate for 1-qubit gates is $0.04\%$ \cite{PhysRevA.77.012307}, while the error for 2-qubit gates is $2.7\%$ \cite{PhysRevLett.125.150505}.

Turning to our main result, the experimental estimation of the mutual information, we see that the mutual information is thus given by $I(A;E) = 0.24311$ (with $95\%$ confidence interval $(0.24279, 0.24344)$). Again, by contrast, we point out that the theoretical bound on the information obtained by Eve is $0.39912$ (see \eqref{Eve_mutual_info_bound}). 

These results agree with what we would expect on the current noisy hardware: errors incurred in the implementation of our cloning circuit mean that the fidelity of the clones is lower than their corresponding theoretical value. As a result, Alice and Bob are able to tolerate more noise in their communication since Eve is inhibited by her inability to successfully clone qubits in a way that agrees with the theoretical calculations for the fidelity. We expect that this gap will close as hardware improves and noise is eliminated from the quantum internet, and note that this will likely be an avenue for future study. Furthermore, as implementations of quantum networks continue to develop, the experiment outlined in this work, together with the statistical analysis developed in Section \ref*{section:stats}, can be used to by legitimate parties to determine the profile of  information obtained by a would-be attacker of the BB84 protocol.





\appendix
\section{Optimization of Phase-Covariant Clones} \label{appendix:optimization}
In this appendix we provide a Lagrange multiplier argument for the optimization of the phase-covariant clones given in Section \ref{clone_optimization}. We recall that the problem of optimizing the clones could be expressed as a constrained optimization problem: 
\begin{equation}
\label{constrained_optimization_problem}
\text{maximize} \quad \eta_{B} = 2 \mu \xi \quad \text{given that} \quad 2 \mu \nu = \eta_{A} \quad \text{and} \quad \mu^{2} + \nu^{2} + \xi^{2} = 1.
\end{equation}
For clarity we emphasize that $\eta_{A}$ is viewed as a fixed constant in \eqref{constrained_optimization_problem}.

To solve the constrained optimization problem \eqref{constrained_optimization_problem} we use the method of Lagrange multipliers: we seek $\lambda_{1}, \lambda_{2}, \mu, \nu, \xi$ satisfying
\begin{equation}
\label{lagrange_equations}
\nabla \eta_{B} = \lambda_{1} \nabla g_{1} + \lambda_{2} \nabla g_{2}, \qquad g_{1}(\mu, \nu, \xi) = 0, \qquad g_{2}(\mu, \nu, \xi) = 0,
\end{equation}
where
\begin{equation*}
g_{1}(\mu, \nu, \xi) = 2 \mu \nu - \eta_{A} \qquad \text{and} \qquad g_{2}(\mu, \nu, \xi) = \mu^{2} + \nu^{2} + \xi^{2} - 1.
\end{equation*}
We rewrite the equations \eqref{lagrange_equations} as a system of (nonlinear) equations:
\begin{align}
\xi &= \lambda_{1} \nu + \lambda_{2} \mu \label{lagrange1}\\
0 &= \lambda_{1} \mu + \lambda_{2} \nu \label{lagrange2}\\
\mu &= \lambda_{2} \xi \label{lagrange3}\\
2 \mu \nu &= \eta_{A} \label{lagrange4}\\
\mu^{2} + \nu^{2} + \xi^{2}& = 1. \label{lagrange5}
\end{align}
We substitute \eqref{lagrange3} into \eqref{lagrange2} to see that $\nu = -\lambda_{1} \xi$.\footnote{Observe that if $\lambda_{2} = 0$ then $\mu = 0$ and it follows that $\eta_{A} = 0$.}
 Inserting these new relationships into \eqref{lagrange1} yields $\lambda_{2}^{2} - \lambda_{1}^{2} = 1$. Note that we have canceled factors of $\xi$ in this last step; if $\xi=0$ then we find that $\eta_{A} = 0$. Further, using \eqref{lagrange5}, we find that
\begin{equation}
\label{opteq1}
\lambda_{2}^{2} \xi^{2} + \lambda_{1}^{2} \xi^{2} + \xi^{2} = 1.
\end{equation}
Since $\lambda_{2}^{2} - \lambda_{1}^{2} = 1$, we can solve \eqref{opteq1} for $\xi$:
\begin{equation*}
\xi = \pm \frac{1}{\sqrt{2} \lambda_{2}}.
\end{equation*}
Using \eqref{lagrange3} we thus have $\mu = \pm \frac{1}{\sqrt{2}}$, and so the constraint equation \eqref{lagrange4} yields $\nu = \pm \frac{\eta_{A}}{\sqrt{2}}$. We now return to the normalization condition \eqref{lagrange5} to see that
\begin{equation*}
	\frac{1}{2} + \frac{1}{2} \eta_{A}^{2} + \xi^{2} = 1.
\end{equation*}	
This means that
\begin{equation*}
\xi = \pm \sqrt{\frac{1 - \eta_{A}^{2}}{2}}.
\end{equation*}
These calculations indicate that the optimal values of $\eta_{B}$ are given by
\begin{equation*}
\eta_{B} = 2\mu \xi = \pm \sqrt{1 - \eta_{A}^{2}}.
\end{equation*}	
In particular, in the case that phase-covariant clones are optimal, the corresponding shrinking factors $\eta_{A}, \eta_{B}$ satisfy the circle relation \eqref{circle_relation}.

\bibliographystyle{plain}
\bibliography{reflist}

\end{document}